\documentclass[twocolumn,times,nonlinenumbers]{aastex631}
\usepackage{graphicx, amsmath, xcolor}
\let\tablenum\relax
\def\edit1#1{#1}
\usepackage{siunitx}
\DeclareSIUnit\gauss{G}
\bibpunct{(}{)}{;}{a}{}{,}

\newcommand{\narrowfig}[3]{%
\begin{figure}[tbp]
\begin{center}
\includegraphics[width=86mm]{#1}
\caption{#3}
\label{#2}
\end{center}
\end{figure}
}

\newcommand{\widefig}[3]{%
\begin{figure*}[tbp]
\begin{center}
\includegraphics[width=180mm]{#1}
\caption{#3}
\label{#2}
\end{center}
\end{figure*}
}

\begin{document}

\title{Derivation of Instrument Requirements for Polarimetry using Mg, Fe, and Mn lines between 250 and 290 nm}
\shorttitle{Derivation of Instrument Requirements for Polarimetry}

\author[0000-0002-5084-4661]{A.G.~de~Wijn}
\author[0000-0001-5174-0568]{P.G.~Judge}
\affil{High Altitude Observatory, National Center for Atmospheric Research, P.O. Box 3000, Boulder, CO 80307, USA}

\author[0000-0002-8504-8470]{R.~Ezzeddine}
\affil{Department of Astronomy, University of Florida, 211 Bryant Space Sciences Center, Gainesville, Florida, 32611, USA}

\author[0000-0002-3234-3070]{A.~Sainz~Dalda}
\affil{Lockheed Martin Solar \& Astrophysics Laboratory, 3251 Hanover Street, Palo Alto, CA 94304, USA}
\affil{Bay Area Environmental Research Institute, NASA Research Park, Moffett Field, CA 94035, USA.}

\correspondingauthor{A.G.~de~Wijn}
\email{dwijn@ucar.edu}

\shortauthors{De Wijn et al.}

\begin{abstract}
Judge et al.\ (2021)\nocite{2021ApJ...917...27J} recently argued that a region of the solar spectrum in the near-UV between about \num{250} and \SI{290}{\nm} is optimal for studying magnetism in the solar chromosphere due to an abundance of \ion{Mg}{2}, \ion{Fe}{2}, and \ion{Fe}{1} lines that sample various heights in the solar atmosphere.
In this paper we derive requirements for spectropolarimetric instruments to observe these lines.
We derive a relationship between the desired sensitivity to magnetic field and the signal-to-noise of the measurement from the weak-field approximation of the Zeeman effect.
We find that many lines will exhibit observable polarization signals for both longitudinal and transverse magnetic field with reasonable amplitudes.
\end{abstract}


\section{Introduction}

The region between \num{250} and \SI{290}{\nm} of the solar spectrum contains the well-known \ion{Mg}{2}~h and~k lines but also a large number of \ion{Fe}{1} and \ion{Fe}{2} lines.
In particular, there are many strong \ion{Fe}{2} lines that sample various heights in the solar atmosphere.
\cite{2021ApJ...917...27J} argue that these lines are highly promising for studying magnetism in the solar chromosphere based on a broad evaluation of possible diagnostics of magnetic field in the solar chromosphere.
In order to evaluate the practical use of this region of the spectrum, they analyzed the signal levels expected in the \ion{Mg}{2}~k line by synthesizing polarized spectra using the HanleRT code \citep{2020ApJ...891...91D} over a grid of line-of-sight angles and magnetic field strengths for a given field inclination and azimuth angle.
They find that the combined Hanle and Zeeman effects produce measurable signals for this line in many geometries.
In particular, they note that diagnostics of vector field with strengths of \num{5} to \SI{50}{\gauss} are achievable for \edit1{observation} angles greater than \ang{45}, as a result of the Hanle effect in Stokes $Q$ and $U$.

However, they do not evaluate \emph{observability} of magnetic field diagnostics using the many \ion{Fe}{2} lines, which underpin the value of this region of the spectrum for diagnostics of magnetic field in the chromosphere.
In this paper we evaluate the signal-to-noise ratio (SNR) required to observe a signature of longitudinal or transverse magnetic field with a given field strength under the assumption of the weak-field approximation of the Zeeman effect.
We do not treat the Hanle effect in our study.
\cite{2021ApJ...917...27J} note that work on the Hanle effect in the \ion{Fe}{2} lines is underway, and will be reported elsewhere.
However, the Stokes $V$ signal is unaffected by the Hanle effect, and therefore the signal strength of the longitudinal component of the magnetic field can be estimated using the methods in this paper.
In addition, many science cases will require observations on the disk or of magnetic field with strengths considerably exceeding the critical Hanle field strength.
In those situations, the Hanle effect has a negligible contribution to the polarization signal, and therefore the methods used in this paper apply.

We will first discuss briefly the weak-field approximation and its applicability, and then derive formulae to relate the error on a measurement of the magnetic field to the SNR of an observation for both the longitudinal and transverse components.
We will derive parameters in several different ways to illustrate possible ways one may approach a similar problem for other spectral lines.
Finally, for specific lines in this particular spectral region, we will investigate the instrumental effect of limited spectral resolution, and illustrate the method through an example calculation.

\section{Analysis}\label{sec:analysis}

The Zeeman splitting of a spectral line is given by
\begin{equation}\label{eq:deltalambdab}
	\Delta\lambda_\mathrm{B}=s\,B\,\lambda_0^2,
\end{equation}
where $B$ is the magnetic field strength and $\lambda_0$ is the rest wavelength of the line.
We here work with vacuum wavelengths in units of $\mathrm{nm}$ and the magnetic flux density $B$ in units of \si{\gauss}, and therefore have $s=\SI{4.67e-11}{\per\nm\per\gauss}$.

If the Zeeman splitting is much smaller than the Doppler width of the line, it is possible to apply a perturbative scheme to the radiative transfer equations and derive expressions for the circular and linear polarization signals in terms of the first and second derivative of the intensity profile, respectively \citep{1973SoPh...31..299L}.
\cite{2004ASSL..307.....L} note that for iron lines in the visible spectrum that sample the photosphere, this approximation is valid up to \si{k\gauss} field strengths.
At shorter wavelengths, the weak-field approximation is applicable for stronger field strengths because Zeeman splitting scales with the square of the wavelength, while Doppler broadening scales linearly.
In addition, lines that form in the chromosphere have larger Doppler broadening than those that form in the photosphere.
For these reasons we can safely assume the weak-field approximation is applicable for magnetic field up to at least \si{k\gauss} strength for the lines we consider here.

The circular and linear polarization signals as a function of wavelength are
\begin{align}
	\label{eq:Vlambda}
	V(\lambda)&=-\Delta\lambda_\mathrm{B}\, \bar g\,\frac{\partial I(\lambda)}{\partial\lambda}\,\cos\theta,\\
	\label{eq:Llambda}
	L(\lambda)&=-\frac14\Delta\lambda_\mathrm{B}^2\, \bar G\, \frac{\partial^2 I(\lambda)}{\partial\lambda^2}\,\sin^2\theta,
\end{align}
where $\bar g$ and $\bar G$ are the effective Land\'e factors for longitudinal and transverse magnetic field, respectively, $I(\lambda)$ is the intensity, and $\theta$ is the inclination of the magnetic field with respect to the line-of-sight.
For convenience, we write $B_\parallel=B\,\cos\theta$ and $B_\bot=B\,\sin\theta$.

An interpretation of the Stokes profiles effectively combines the signal over a wavelength range $\Delta\lambda$.
We are therefore interested in the integrated absolute signals $\mathcal{V}$ and $\mathcal{L}$,
\begin{align}
	\mathcal{V}&=\int_{\Delta\lambda}\left|V(\lambda)\right|\nonumber\\
			   &=s\,\bar g\,\lambda_0^2\,\left|B_\parallel\right|\,\int_{\Delta\lambda}\left|\frac{\partial I(\lambda)}{\partial\lambda}\right|\,\mathrm{d}\lambda\label{eq:VV},\\
	\mathcal{L}&=\frac14\,s^2\,\bar G\,\lambda_0^4\,B_\bot^2\,\int_{\Delta\lambda}\left|\frac{\partial^2 I(\lambda)}{\partial\lambda^2}\right|\,\mathrm{d}\lambda\label{eq:LL}.
\end{align}

Our goal is to evaluate the expected errors $\sigma_{B\parallel}$ on $B_\parallel$ and $\sigma_{B\bot}$ on $B_\bot$ in some way that can be readily estimated, such as a function of the SNR of the intensity measurement that can be determined from a flux budget calculation.
We therefore also define the integrated intensity signal,
\begin{equation}\label{eq:II}
	\mathcal{I}=\int_{\Delta\lambda}I(\lambda)\,\mathrm{d}\lambda,
\end{equation}
and note that the uncertainties $\sigma_\mathcal{V}$ of $\mathcal{V}$ and $\sigma_\mathcal{L}$ of $\mathcal{L}$ are related to the uncertainty $\sigma_\mathcal{I}$ of $\mathcal{I}$ through the modulation efficiencies $\epsilon_I$, $\epsilon_Q$, $\epsilon_U$, and $\epsilon_V$ in Stokes $I$, $Q$, $U$, and $V$ \citep{2000ApOpt..39.1637D},
\begin{align}
	\sigma_\mathcal{V}&=\frac{\epsilon_I}{\epsilon_V}\,\sigma_\mathcal{I},\label{eq:sigmaVV}\\
	\sigma_\mathcal{L}&=\frac{\epsilon_I}{\epsilon_L}\,\sigma_\mathcal{I},\label{eq:sigmaLL}
\end{align}
where we have written $\epsilon_L$ for the modulation efficiency of the linear polarization signal of interest.

Since Eq.~\ref{eq:VV} relates $\mathcal{V}$ to $B_\parallel$, we can express the uncertainty $\sigma_{B\parallel}$ as the uncertainty of $\mathcal{V}$, and subsequently of $\mathcal{I}$ by substituting Eq.~\ref{eq:sigmaVV},
\begin{align}
	\sigma_{B\parallel}&=\frac1{s\,\bar g\,\lambda_0^2}\,\left(\int_{\Delta\lambda}\left|\frac{\partial I(\lambda)}{\partial\lambda}\right|\,\mathrm{d}\lambda\right)^{-1}\,\sigma_\mathcal{V}\\
					   &=\frac1{s\,\bar g\,\lambda_0^2}\,\frac{\epsilon_I}{\epsilon_V}\,\left(\int_{\Delta\lambda}\left|\frac{\partial I(\lambda)}{\partial\lambda}\right|\,\mathrm{d}\lambda\right)^{-1}\,\sigma_\mathcal{I}.\label{eq:sigmaBparsigmaI}
\end{align}
\edit1{Note that we have used the property of the WFA that $I$ does not depend on $B$.}
If we can express the integral of $|\partial I(\lambda)/\partial\lambda|$ in terms of $\mathcal{I}$, then $\sigma_{B\parallel}$ can be expressed in the SNR of the intensity measurement, $\mathcal{I}/\sigma_\mathcal{I}$.

Equivalently, propagating the error in $B_\bot$ through Eq.~\ref{eq:LL} and substituting Eq.~\ref{eq:sigmaLL}, we find
\begin{align}
	\sigma_{B\bot}&=\frac2{s^2\,\bar G\,\lambda_0^4}\,\frac1{B_\bot}\,\left(\int_{\Delta\lambda}\left|\frac{\partial^2 I(\lambda)}{\partial\lambda^2}\right|\,\mathrm{d}\lambda\right)^{-1}\,\sigma_\mathcal{L}\\
				  &=\frac2{s^2\,\bar G\,\lambda_0^4}\,\frac1{B_\bot}\,\frac{\epsilon_I}{\epsilon_P}\,\left(\int_{\Delta\lambda}\left|\frac{\partial^2 I(\lambda)}{\partial\lambda^2}\right|\,\mathrm{d}\lambda\right)^{-1}\,\sigma_\mathcal{I}\label{eq:sigmaBbotsigmaI}.
\end{align}
In this case, we want to express the integral of $|\partial^2 I/\partial\lambda^2|$ in terms of $\mathcal{I}$.
We note, however, that $\sigma_{B\bot}$ is a function of $B_\bot$, and therefore the same SNR in $\mathcal{I}$ will yield a different measurement error depending on the strength of the field being measured.
Notably, $\sigma_{B\bot}$ is infinite for $B_\bot=0$.

In practice, $\mathcal{I}$ will be the sum of a series of discrete measurements.
Each pixel samples the signal weighted with some point spread function, which causes cancellation of some amount of signal.
This effect is discussed in more detail in Sect.~\ref{sec:ie}.
We assume here that the instrument is a spectrograph that samples the spectrum critically with a resolution $R$.
However, the analysis for a different type of instrument, such as a wavelength-tunable imager, is analogous.
The number of measurements $N$ that spans the wavelength range $\Delta\lambda$ is given by
\begin{equation}\label{eq:NfromR}
	N=\frac{2R\,\Delta\lambda}{\lambda_0}.
\end{equation}

There are several ways to approach expressing the first and second derivatives of $I(\lambda)$ in terms of itself.
We will examine a simple approximation of the gradient in terms of the intensity and a characteristic wavelength interval, and a numeric calculation from simulated data or from observations of the intensity spectrum.

\begin{figure*}[tbp]
\begin{center}
\includegraphics[width=180mm]{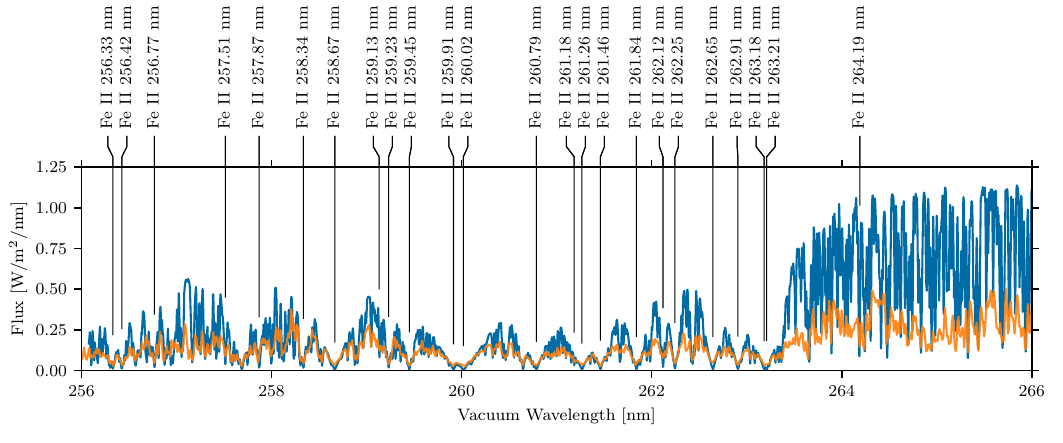}
\includegraphics[width=180mm]{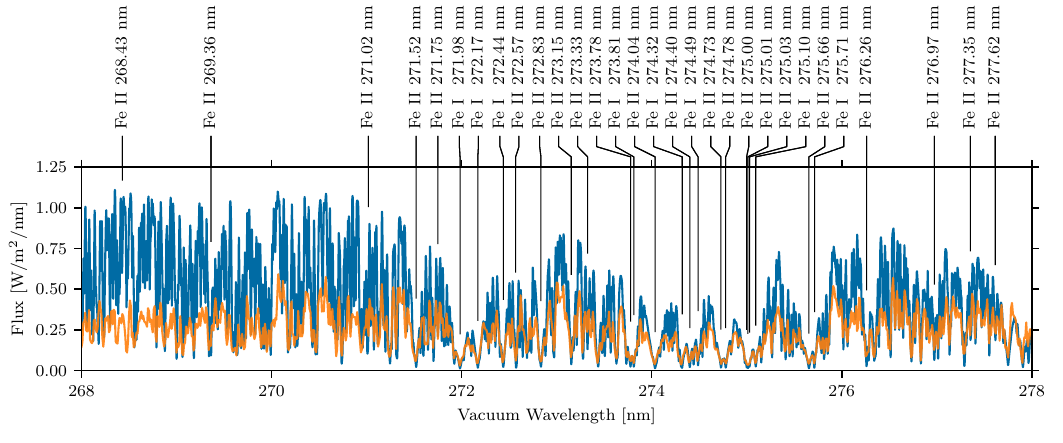}
\caption{Blue lines: synthetic spectrum in two NUV windows that contain \ion{Fe}{1} and \ion{Fe}{2} lines of interest.
Orange lines: measured flux from the 1983 Air Force Geophysics Laboratory (AFGL) balloon measurement.
The synthetic spectrum shows very good agreement with the measured spectrum for the \ion{Fe}{1} and \ion{Fe}{2} lines.}
\label{fig:synspec}
\end{center}
\end{figure*}

\subsection{Characteristic Wavelength Interval}\label{sec:cwi}

This simple approximation equates the gradient as the ratio of the intensity and some characteristic wavelength interval,
\begin{equation}\label{eq:Ityp}
	\left|\frac{\partial I(\lambda)}{\partial\lambda}\right|\approx\frac{I(\lambda)}{\Delta\lambda}.
\end{equation}
Substituting Eq.~\ref{eq:Ityp} in Eq.~\ref{eq:sigmaBparsigmaI}, we find
\begin{equation}
	\sigma_{B\parallel}=\frac{\Delta\lambda}{s\,\bar g\,\lambda_0^2}\,\frac{\epsilon_I}{\epsilon_V}\,\frac{\sigma_\mathcal{I}}{\mathcal{I}}.\label{eq:sigmaBparallel}
\end{equation}
We approximate the pixel SNR,
\begin{equation}\label{eq:isigmaiapprox}
	I\approx\frac1N\,\mathcal{I},\quad
	\sigma_I\approx\frac1{\sqrt{N}}\,\sigma_\mathcal{I}.
\end{equation}
If the measurement noise is dominated by photon statistics, as is often the case, that property is preserved also for the average intensity $I$, i.e., $\sigma_I$ is approximately the square root of $I$, and therefore the SNR $I/\sigma_I$ is the average SNR of the measurement over the wavelength range $\Delta\lambda$.
Substitution of Eqs.~\ref{eq:isigmaiapprox} and~\ref{eq:NfromR} in Eq.~\ref{eq:sigmaBparallel} yields
\begin{equation}
	\sigma_{B\parallel}=\frac1{s\,\bar g\,\lambda_0}\,\sqrt{\frac{\Delta\lambda}{2R\,\lambda_0}}\,\frac{\epsilon_I}{\epsilon_V}\,\frac{\sigma_I}{I}.
\end{equation}

We apply this method to estimate the sensitivity of the \ion{Mg}{2}~h line.
For this line, we have $\bar g=\num{1.33}$, and $\lambda_0=\SI{280}{\nm}$.
We estimate $\Delta\lambda=\SI{0.03}{\nm}$ (equivalent to a velocity of \SI{\pm15}{\km\per\s}, see Eq.~\ref{eq:deltalambdav}).
We now find
\begin{equation}
	\sigma_{B\parallel}=\num{7.58e5}\,\frac1{\sqrt{R}}\,\frac{\sigma_I}{I}~\si{\gauss}.
\end{equation}
For example, if we evaluate this equation for a instrument with $R=\num{30000}$ and require $\sigma_{B\parallel}\le\SI{8}{\gauss}$, we find that the SNR in $I$ must be at least \num{547}.

\subsection{Observed or Simulated Intensity Spectra}\label{sec:osis}

The estimation in Sect.~\ref{sec:cwi} is obviously crude as it relies on a good estimate of $\Delta\lambda$, which is difficult without some prior knowledge of the line profile.
A much better estimate can be derived from intensity spectra that were observed or computed with a numerical radiative transfer code such as RH \citep{2001ApJ...557..389U} or TURBOSPECTRUM \citep{1998A&A...330.1109A,2012ascl.soft05004P} using a model atmosphere.
We can numerically calculate the integral of the absolute value of the gradient as a fraction of $\mathcal{I}$,
\begin{equation}\label{eq:gamma}
	\gamma=\frac1{\mathcal{I}}\,
	\int_{\Delta\lambda}\left|\frac{\partial I(\lambda)}{\partial\lambda}\right|\,\mathrm{d}\lambda,
\end{equation}
Using again Eqs.~\ref{eq:isigmaiapprox} and~\ref{eq:NfromR}, we now have
\begin{align}
	\sigma_{B\parallel}&=\frac1{s\,\bar g\,\lambda_0^2}\,\frac1\gamma\,\sqrt{\frac{\lambda_0}{2\Delta\lambda}}\,\frac{\epsilon_I}{\sqrt{R}\,\epsilon_V}\,\frac{\sigma_I}{I}.
\end{align}
We now define the sensitivity factor for the error on the longitudinal field,
\begin{equation}\label{eq:biggamma}
	\Gamma=\gamma\,s\,\bar g\,\lambda_0^2\,\sqrt{\frac{2\Delta\lambda}{\lambda_0}}
\end{equation}
that captures the properties of a spectral line.
Lines with larger $\Gamma$ have higher sensitivity to longitudinal field, i.e., a requirement for a particular sensitivity of $B_\parallel$ can be met with a measurement with lower SNR.

Similarly for the transverse component of the magnetic field we can numerically calculate the integral of the second derivative,
\begin{equation}\label{eq:chi}
	\chi=\frac1{\mathcal{I}}\,
	\int_{\Delta\lambda}\left|\frac{\partial^2 I(\lambda)}{\partial\lambda^2}\right|\,\mathrm{d}\lambda,
\end{equation}
and find
\begin{equation}
	\sigma_{B\bot}=\frac1{s^2\,\bar G\,\lambda_0^4}\,\frac1\chi\,\sqrt{\frac{2\lambda_0}{\Delta\lambda}}\,\frac1{B_\bot}\,\frac{\epsilon_I}{\sqrt{R}\,\epsilon_L}\,\frac{\sigma_I}{I}.
\end{equation}
Therefore, we define the sensitivity factor for the error on the transverse field,
\begin{equation}\label{eq:bigchi}
	\mathrm{X}=\chi\,s^2\,\bar G\,\lambda_0^4\,\sqrt{\frac{\Delta\lambda}{2\lambda_0}}.
\end{equation}

We use the above procedure to calculate $\Gamma$ and $\mathrm{X}$ for the lines listed in \cite{2021ApJ...917...27J} from a synthetic spectrum and observations from the IRIS mission \citep{2014SoPh..289.2733D}.
Figure~\ref{fig:synspec} shows the synthetic spectrum calculated with TURBOSPECTRUM from a custom Sun-like MARCS model atmosphere \citep{2008A&A...486..951G} and using a line list adopted from the VALD database \citep{1995A&AS..112..525P,2015PhyS...90e4005R}.
This spectrum was calculated under the assumption of local thermal equilibrium \edit1{(LTE)}, but shows very good agreement with measured spectra for the \ion{Fe}{1} and \ion{Fe}{2} lines, such as the AFGL balloon measurements \citep{1991JGR....9612927H} shown also in Fig.~\ref{fig:synspec}.
\edit1{The synthesis includes \num{102691} lines from \num{115} atomic and \num{10} molecular species between \num{256} and \SI{285}{\nm}.}

\edit1{The MARCS model atmosphere does not include a chromospheric temperature rise.
However, using a more realistic atmospheric model does not necessarily result in a more realistic synthetic spectrum.
A model atmosphere like FAL-C \citep{1993ApJ...406..319F} would result in emission peaks in the cores of strong lines that are not observed in the AFGL spectra.
These peaks are the result of the assumption of LTE that is not valid in the line cores, and the peaks would not be present if non-LTE physics (e.g., scattering and partial redistribution) were included in the spectral synthesis.}

\edit1{We can expect the intensity in the cores of strong lines to saturate in this synthesis (see, e.g., the upper-left panel of Fig.~\ref{fig:tblines}).
The resultant line core profile is nearly flat and exhibits only a small gradient and second derivative, and hence little circular or linear polarization signal is produced in the presence of magnetic field.
Therefore, the analysis presented here based on this synthesis will produce lower values of $\Gamma$ and $\mathrm{X}$ than one based on a synthesis that incorporates pertinent non-LTE effects and uses a more realistic atmospheric model.}

IRIS observes the solar spectrum around the \ion{Mg}{2} h and k lines that also includes the \ion{Mn}{1} lines used by \cite{2021SciA....7.8406I} to infer longitudinal magnetic field from data from the CLASP2 flight \citep{2016SPIE.9905E..08N,2020SPIE11444E..6WT}.
We choose a data set of NOAA AR12957 taken on March~4, 2022 around 10~UT.
This observation contains a region of plage, the edge of a sunspot, and some more quiet areas.
The left panel of Fig.~\ref{fig:mgIIk_factors} shows the intensity of the \ion{Mg}{2}~k core.
We could compute the longitudinal and transverse sensitivity factors for every pixel in the map.
However, we would overestimate the factors due to measurement noise that creates spurious signals in first and second derivative of the intensity profile.
We therefore use ``representative profiles'' (RPs) from the IRIS$^2$ database \citep{2019ApJ...875L..18S}.
An RP is an average of many similar line profiles, and therefore has very low noise, so that we can accurately determine the first and second derivative of the intensity.
There are 160 RPs in this map.
More than half the pixels are represented by the most popular 25 RPs, and only 3 RPs represent less than 100 pixels each.

We note that we can express $\Delta\lambda$ in terms of a velocity $v$,
\begin{equation}\label{eq:deltalambdav}
	\Delta\lambda=2\,\frac{v}{c}\lambda_0.
\end{equation}
We use $v=\SI{12.5}{\km\per\s}$ for the \ion{Fe}{1}, \ion{Fe}{2}, and \ion{Mg}{2} lines, and $v=\SI{7.5}{\km\per\s}$ for the \ion{Mn}{1} lines.
These velocities give reasonable integration intervals and more or less correspond to typical sound speed estimates for the chromosphere and photosphere.
The values for $\Gamma$ and $\mathrm{X}$ are not strongly dependent on the choice of $v$, since the wings of the lines tend not to contribute significant polarization signal (see Fig.~\ref{fig:tblines}).
However, other nearby spectral lines may contribute spurious polarization signal in the wavelength window.
We therefore limit the window to the nearest local maximum of the intensity spectrum to reduce contamination by other lines (see Fig.~\ref{fig:tblines} panels in the middle row left and center column).

\startlongtable
\begin{deluxetable*}{RRRlR|RR|RR|l}
\tablecaption{\label{tab:lines}%
Longitudinal and transverse field sensitivity factors for prominent lines in the  solar chromospheric spectrum.
$\Delta\lambda$ denotes the integration window.
Land\'e g-factors are computed using LS coupling.
Wavelength in air is included for easy reference against Table~1 in \cite{2021ApJ...917...27J}.}%
\tablehead{%
\colhead{$\lambda_0$} & \colhead{$\lambda_{0,\mathrm{air}}$} & \colhead{$\Delta\lambda$} &
\colhead{Ion} & \multicolumn{1}{c|}{$\log\tau_0$} &
\colhead{$g$} & \multicolumn{1}{c|}{$\Gamma$} &
\colhead{$G$} & \multicolumn{1}{c|}{$\mathrm{X}$} &
\colhead{Blend} \\[-2mm]
\colhead{(\si{\nm})} & \colhead{(\si{\nm})} & \colhead{(\si{\pm})} & & &
& \multicolumn{1}{c|}{(\SI{e-5}{\per\gauss})} &
& \multicolumn{1}{c|}{(\SI{e-8}{\per\gauss\squared})} & }
\startdata
256.331 & 256.408 & \num{-11}/\num[retain-explicit-plus]{ +7} & \ion{Fe}{2} & -2.54 & 1.21 & 1.17 & 1.46 & \num{2.45} & minor \\
256.425 & 256.502 & \num{ -5}/\num[retain-explicit-plus]{ +8} & \ion{Fe}{2} & -2.87 & 1.10 & 1.21 & 1.18 & \num{2.86} & major \\
256.768 & 256.845 & \num{\pm  11} & \ion{Fe}{2} & -3.30 & 0.83 & 0.94 & 0.56 & \num{1.15} &  \\
257.514 & 257.591 & \num{\pm  11} & \ion{Fe}{2} & -4.33 & 1.30 & 1.77 & 1.65 & \num{3.03} & minor \\
257.869 & 257.946 & \num{-11}/\num[retain-explicit-plus]{ +6} & \ion{Fe}{2} & -3.32 & 1.33 & 1.53 & -0.01 & \num{0.01} & major \\
258.335 & 258.413 & \num{\pm  11} & \ion{Fe}{2} & -3.19 & 1.47 & 1.62 & 1.74 & \num{3.59} &  \\
258.665 & 258.743 & \num{\pm  11} & \ion{Fe}{2} & -2.03 & 1.50 & 0.73 & 2.25 & \num{1.00} & minor \\
259.132 & 259.210 & \num{-11}/\num[retain-explicit-plus]{+10} & \ion{Fe}{2} & -4.47 & 1.50 & 1.71 & 2.25 & \num{4.27} & severe \\
259.232 & 259.309 & \num{ -8}/\num[retain-explicit-plus]{+11} & \ion{Fe}{2} & -3.24 & 1.49 & 1.61 & 2.03 & \num{4.08} & minor \\
259.451 & 259.528 & \num{ -8}/\num[retain-explicit-plus]{+11} & \ion{Fe}{2} & -4.02 & 2.17 & 1.80 & 4.52 & \num{7.14} & minor \\
259.915 & 259.992 & \num{\pm  11} & \ion{Fe}{2} & -1.93 & 1.50 & 0.73 & 2.25 & \num{0.75} &  \\
260.018 & 260.095 & \num{\pm  11} & \ion{Fe}{2} & -1.39 & 1.56 & 0.42 & 2.42 & \num{0.30} & minor \\
260.787 & 260.865 & \num{\pm  11} & \ion{Fe}{2} & -1.99 & 1.50 & 0.82 & 2.24 & \num{1.28} & minor \\
261.185 & 261.263 & \num{-11}/\num[retain-explicit-plus]{ +6} & \ion{Fe}{2} & -4.23 & 1.90 & 2.30 & 3.58 & \num{9.01} & minor \\
261.265 & 261.343 & \num{\pm  11} & \ion{Fe}{2} & -1.79 & 1.59 & 0.73 & 2.52 & \num{0.83} &  \\
261.460 & 261.538 & \num{\pm  11} & \ion{Fe}{2} & -2.20 & 1.50 & 0.94 & 2.12 & \num{1.54} &  \\
261.840 & 261.918 & \num{\pm  11} & \ion{Fe}{2} & -2.36 & 1.66 & 1.13 & 2.75 & \num{2.42} &  \\
262.119 & 262.197 & \num{-11}/\num[retain-explicit-plus]{ +5} & \ion{Fe}{2} & -3.75 & 1.87 & 2.31 & 3.49 & \num{9.82} & major \\
262.245 & 262.323 & \num{\pm  11} & \ion{Fe}{2} & -2.76 & 3.34 & 3.10 & 11.15 & \num{17.84} &  \\
262.645 & 262.724 & \num{\pm  11} & \ion{Fe}{2} & -2.08 & 1.50 & 0.98 & 2.25 & \num{1.75} &  \\
262.907 & 262.986 & \num{\pm  11} & \ion{Fe}{2} & -2.21 & 1.50 & 1.04 & 2.12 & \num{1.79} &  \\
263.183 & 263.262 & \num{-11}/\num[retain-explicit-plus]{+10} & \ion{Fe}{2} & -2.02 & 1.50 & 0.85 & 2.24 & \num{1.44} & major \\
263.210 & 263.289 & \num{\pm  11} & \ion{Fe}{2} & -1.97 & 1.50 & 0.81 & 2.25 & \num{1.41} & major \\
264.191 & 264.269 & \num{ -6}/\num[retain-explicit-plus]{+11} & \ion{Fe}{2} & -4.62 & 1.87 & 2.61 & 3.39 & \num{13.36} & severe \\
268.431 & 268.510 & \num{ -4}/\num[retain-explicit-plus]{ +2} & \ion{Fe}{2} & -4.99 & 1.84 & 0.64 & 3.34 & \num{2.76} & severe \\
269.363 & 269.443 & \num{ -8}/\num[retain-explicit-plus]{+11} & \ion{Fe}{2} & -4.40 & 1.93 & 3.42 & 3.64 & \num{10.36} & major \\
271.018 & 271.099 & \num{ -4}/\num[retain-explicit-plus]{+11} & \ion{Fe}{2} & -4.30 & 2.10 & 3.53 & 4.17 & \num{19.21} & severe \\
271.521 & 271.602 & \num{\pm  11} & \ion{Fe}{2} & -3.09 & 1.50 & 1.53 & 2.25 & \num{3.82} &  \\
271.751 & 271.831 & \num{\pm   5} & \ion{Fe}{2} & -3.11 & 1.33 & 2.83 & 1.55 & \num{8.63} & severe \\
271.984 & 272.064 & \num{\pm  11} & \ion{Fe}{1} & -4.45 & 1.25 & 0.92 & 1.54 & \num{1.77} & minor \\
272.171 & 272.251 & \num{\pm  11} & \ion{Fe}{1} & -4.76 & 1.17 & 1.07 & 1.32 & \num{2.31} &  \\
272.439 & 272.519 & \num{ -8}/\num[retain-explicit-plus]{+11} & \ion{Fe}{1} & -5.22 & 1.00 & 1.08 & 0.85 & \num{1.67} & minor \\
272.569 & 272.649 & \num{-11}/\num[retain-explicit-plus]{ +4} & \ion{Fe}{2} & -3.05 & 1.20 & 1.38 & 1.03 & \num{2.48} & severe \\
272.835 & 272.916 & \num{\pm  11} & \ion{Fe}{2} & -3.05 & 1.50 & 1.47 & 2.25 & \num{4.19} & minor \\
273.154 & 273.235 & \num{\pm  11} & \ion{Fe}{2} & -3.19 & 0.80 & 0.97 & -0.29 & \num{0.64} &  \\
273.326 & 273.407 & \num{ -9}/\num[retain-explicit-plus]{+10} & \ion{Fe}{2} & -4.92 & 1.45 & 2.63 & 1.60 & \num{5.32} & major \\
273.778 & 273.859 & \num{ -6}/\num[retain-explicit-plus]{+11} & \ion{Fe}{2} & -3.22 & 1.50 & 1.61 & 2.16 & \num{4.88} & severe \\
273.812 & 273.893 & \num{\pm  11} & \ion{Fe}{1} & -5.12 & 2.00 & 2.08 & 3.26 & \num{7.55} &  \\
274.036 & 274.117 & \num{\pm  11} & \ion{Fe}{2} & -3.31 & 1.43 & 1.05 & 2.04 & \num{2.02} & minor \\
274.322 & 274.403 & \num{ -5}/\num[retain-explicit-plus]{+11} & \ion{Fe}{1} & -5.01 & 1.67 & 1.55 & 2.52 & \num{5.56} & severe \\
274.401 & 274.482 & \num{\pm  11} & \ion{Fe}{2} & -2.81 & 0.50 & 0.47 & 0.24 & \num{0.39} &  \\
274.488 & 274.569 & \num{\pm  11} & \ion{Fe}{1} & -5.51 & 2.50 & 2.78 & 6.27 & \num{15.09} &  \\
274.729 & 274.810 & \num{\pm  11} & \ion{Fe}{2} & -2.59 & 0.90 & 0.77 & 0.80 & \num{1.15} &  \\
274.779 & 274.861 & \num{\pm  11} & \ion{Fe}{2} & -2.66 & 1.37 & 1.17 & 1.88 & \num{2.54} &  \\
274.999 & 275.081 & \num{-11}/\num[retain-explicit-plus]{ +6} & \ion{Fe}{2} & -3.02 & 1.20 & 1.03 & 1.44 & \num{2.84} & severe \\
275.013 & 275.095 & \num{ -8}/\num[retain-explicit-plus]{+11} & \ion{Fe}{2} & -2.38 & 1.07 & 0.80 & 1.14 & \num{1.69} & severe \\
275.030 & 275.112 & \num{ -5}/\num[retain-explicit-plus]{+11} & \ion{Fe}{2} & -3.24 & 0.00 & 0.00 & 0.00 & \num{0.00} & severe \\
275.095 & 275.177 & \num{\pm  11} & \ion{Fe}{1} & -5.04 & 1.58 & 1.65 & 2.38 & \num{5.64} & minor \\
275.655 & 275.737 & \num{\pm  11} & \ion{Fe}{2} & -2.17 & 1.17 & 0.85 & 1.35 & \num{1.23} &  \\
275.714 & 275.796 & \num{ -4}/\num[retain-explicit-plus]{+11} & \ion{Fe}{1} & -5.52 & 2.00 & 1.71 & 3.99 & \num{8.45} & severe \\
276.263 & 276.344 & \num{-12}/\num[retain-explicit-plus]{ +7} & \ion{Fe}{2} & -3.25 & 1.50 & 1.89 & 2.16 & \num{5.01} & major \\
276.975 & 277.057 & \num{ -7}/\num[retain-explicit-plus]{+12} & \ion{Fe}{2} & -3.08 & 1.50 & 2.26 & 2.25 & \num{5.84} & severe \\
277.355 & 277.437 & \num{ -6}/\num[retain-explicit-plus]{ +5} & \ion{Fe}{2} & -3.14 & 1.50 & 2.86 & 2.25 & \num{9.38} & severe \\
277.616 & 277.698 & \num{ -6}/\num[retain-explicit-plus]{ +4} & \ion{Fe}{2} & -6.06 & 1.34 & 2.36 & 0.44 & \num{1.84} & severe \\
279.564 & 279.647 & \num{\pm   7} & \ion{Mn}{1} & -5.65 & 1.98 & 0.64 & 1.80 & \num{0.42} & minor \\
279.635 & 279.718 & \num{\pm  12} & \ion{Mg}{2} & 0.00 & 1.17 & 0.27 & 1.33 & \num{0.13} &  \\
279.909 & 279.992 & \num{\pm   7} & \ion{Mn}{1} & -5.79 & 1.70 & 1.01 & 3.74 & \num{1.44} &  \\
280.191 & 280.273 & \num{\pm   7} & \ion{Mn}{1} & -5.96 & 0.84 & 0.48 & 2.85 & \num{1.09} &  \\
280.353 & 280.435 & \num{\pm  12} & \ion{Mg}{2} & -0.30 & 1.33 & 0.31 & 1.33 & \num{0.15} &  \\
\enddata
\end{deluxetable*}

Results are given in Table~\ref{tab:lines} and shown in Figs.~\ref{fig:mgIIk_factors} and~\ref{fig:mgIIhk_cdf}.
The line profiles were visually evaluated and qualitatively categorized as suffering from blends with varying severity given in the ``Blend'' column.
The $\Gamma$ and $\mathrm{X}$ values of a line with a major or severe blend are likely affected and overestimate the true values, as the blends create additional gradients.
The values for the \ion{Mg}{2} and \ion{Mn}{1} lines in the table are derived from the spectrum of the most popular RP that represents \num{10697} pixels (3.2\% of the FOV).
We processed each RP, and map the longitudinal and transverse sensitivity factors back to pixels in the FOV, shown in the center and right panels of Fig.~\ref{fig:mgIIk_factors}, respectively.
Figure~\ref{fig:mgIIhk_cdf} shows the cumulative probability density functions for the longitudinal and transverse sensitivity factors.
The most popular RP is around the 70\% and 85\% percentiles for $\Gamma$ and $\mathrm{X}$, respectively.

\widefig{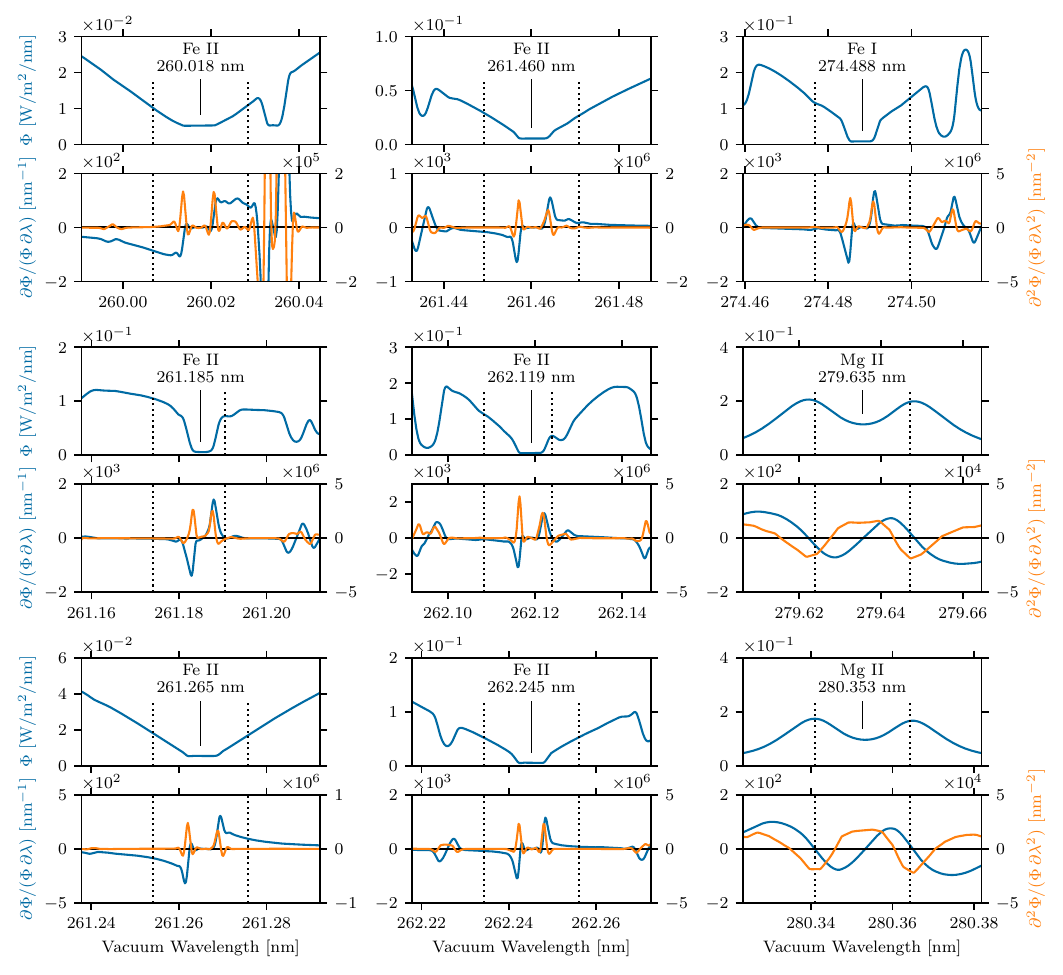}{fig:tblines}{Intensity, and first and second derivative of intensity for selected lines.}

We select a series of \ion{Fe}{2} lines, a single \ion{Fe}{1} line, and the \ion{Mg}{2}~h and~k lines for closer study.
We pick \ion{Fe}{1} and \ion{Fe}{2} lines that are preferentially unaffected by blends, sample \edit1{at} optical depths down to the photosphere in roughly equal steps of $\log\tau$, have large $\Gamma$ and $\mathrm{X}$, and are nearby one another in the spectrum.
This selection procedure results nine lines in two distinct wavelength regions of interest: between \ion{Fe}{2} \SI{260.018}{\nm} and \ion{Fe}{2} \SI{262.245}{\nm}, and between \ion{Fe}{1} \SI{274.488}{\nm} and \ion{Mg}{2}~h \SI{280.353}{\nm}.
Figure~\ref{fig:tblines} shows the synthetic spectra for these lines, together with the first and second derivatives.

\widefig{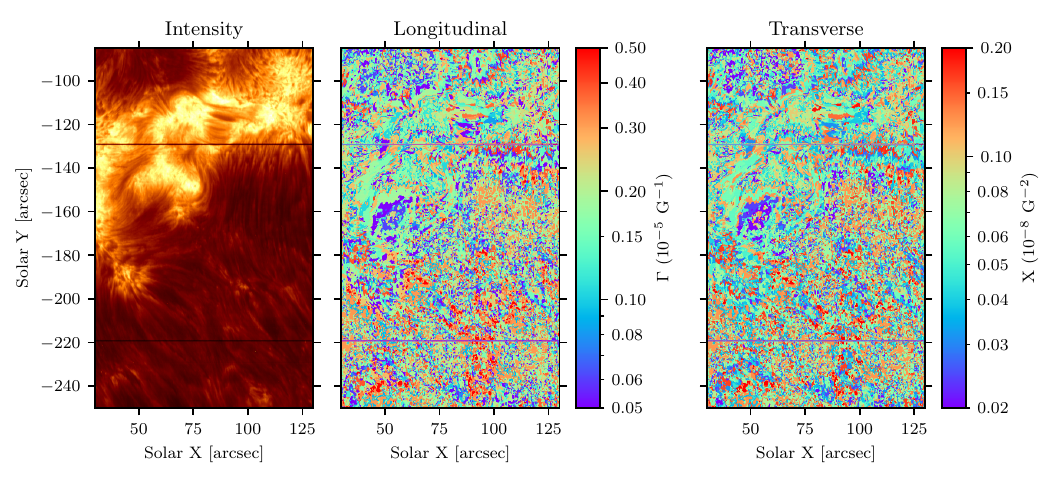}{fig:mgIIk_factors}{%
	IRIS map used to calculate the longitudinal and transverse sensitivity factors for \ion{Mg}{2} and \ion{Mn}{1} lines.
	Left panel: intensity in the core of the \ion{Mg}{2}~k line.
	Center panel: longitudinal field sensitivity factor $\Gamma$.
	Right panel: transverse field sensitivity factor $\mathrm{X}$.}

\narrowfig{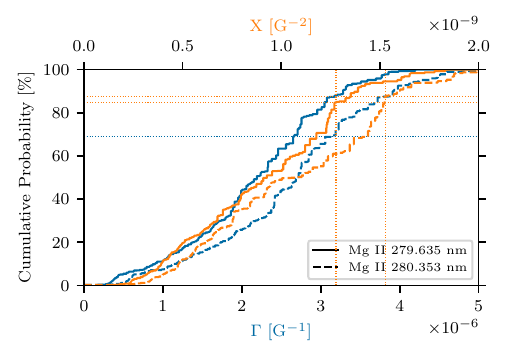}{fig:mgIIhk_cdf}{Cumulative distribution function of the longitudinal and transverse sensitivity factors $\Gamma$ and $\mathrm{X}$ for the \ion{Mg}{2}~h and~k lines.
The $\Gamma$ and $\mathrm{X}$ values and the cumulative probabilities for the most popular RP (also given in Table~\ref{tab:lines}) are indicated by vertical and horizontal dotted lines, respectively.}

\section{Instrumental Effects}\label{sec:ie}

In practice, all measurements will be affected by instrumental effects.
As already mentioned in Sect.~\ref{sec:analysis}, $\mathcal{I}$ will be the sum of a series of discrete measurements that sample the signal weighted with some line spread function (LSF).
We evaluate here how this affects the required measurement sensitivity.
For the sake of simplicity, we assume each measurement is affected by the same LSF $\rho(\lambda)$.
Equations~\ref{eq:II}, \ref{eq:gamma}, and \ref{eq:chi} then become
\begin{align}
	\mathcal{I}'&=\int_{\Delta\lambda}(\rho\ast I)(\lambda)\,\mathrm{d}\lambda,\\
	\gamma'&=\frac1{\mathcal{I}'}\,
	\int_{\Delta\lambda}\left|\left(\rho\ast\frac{\partial I}{\partial\lambda}\right)(\lambda)\right|\,\mathrm{d}\lambda,\\
	\chi'&=\frac1{\mathcal{I}'}\,
	\int_{\Delta\lambda}\left|\left(\rho\ast\frac{\partial^2 I}{\partial\lambda^2}\right)(\lambda)\right|\,\mathrm{d}\lambda,
\end{align}
where $\ast$ denotes convolution.
It is straightforward to implement this in the numerical analysis presented in Sect.~\ref{sec:osis}.

The LSF of a spectrograph depends on its specific configuration \citep{2014JOSAA..31.2002C}.
The LSF of a typical spectrograph operating in Littrow condition that also satisfies the ``pixel-matching'' condition, i.e., the projected width of the slit is equal to the width of a camera pixel, is approximately Gaussian after accounting for sampling.
We therefore choose to model the LSF as a Gaussian function.

Signal loss factors are shown as a function of spectral resolution for a collection of \ion{Fe}{1} and \ion{Fe}{2} lines and the \ion{Mg}{2} lines in Fig.~\ref{fig:SNR_loss}.
Stokes~$V$ signals are less affected than linear polarization signals at reasonable spectral resolution.
Linear polarization signal loss varies considerably from line to line, but generally lines that form deeper in the atmosphere have narrower profiles that require higher spectral resolution to achieve the same loss factor.
The \ion{Fe}{2} \SI{260.018}{\nm} line is affected by a blend that at spectral resolutions below about \num{40000} starts to contaminate the polarization signal, causing a spurious rise in the $L$ signal loss factor.

\section{Example SNR Calculation}\label{sec:example}

We now show an example using the above calculations to derive measurement requirements, i.e., SNR on Stokes $I$, for a hypothetical instrument that observes the nine spectral lines previously selected.

Substituting $\gamma'$ and $\chi'$ for $\gamma$ and $\chi$ in Eqs.~\ref{eq:biggamma} and~\ref{eq:bigchi} to account for instrument spectral resolution yields
\begin{align}
	\Gamma'&=2\,\gamma'\,s\,\bar g\,\lambda_0^2\,\sqrt{\frac{v}{c}}\label{eq:biggammaprime}\\
	\mathrm{X}'&=\chi'\,s^2\,\bar G\,\lambda_0^4\,\sqrt{\frac{v}{c}}.\label{eq:bigchiprime}
\end{align}
The errors on the longitudinal and transverse magnetic field are given by
\begin{align}
	\sigma_{B\parallel}&=\frac1{\Gamma'}\,\frac{\epsilon_I}{\sqrt{R}\,\epsilon_V}\,\frac{\sigma_I}{I},\label{eq:sigmabparallel}\\
	\sigma_{B\bot}&=\frac1{\mathrm{X}'}\,\frac1{B_\bot}\,\frac{\epsilon_I}{\sqrt{R}\,\epsilon_L}\,\frac{\sigma_I}{I}.\label{eq:sigmabbot}
\end{align}

\narrowfig{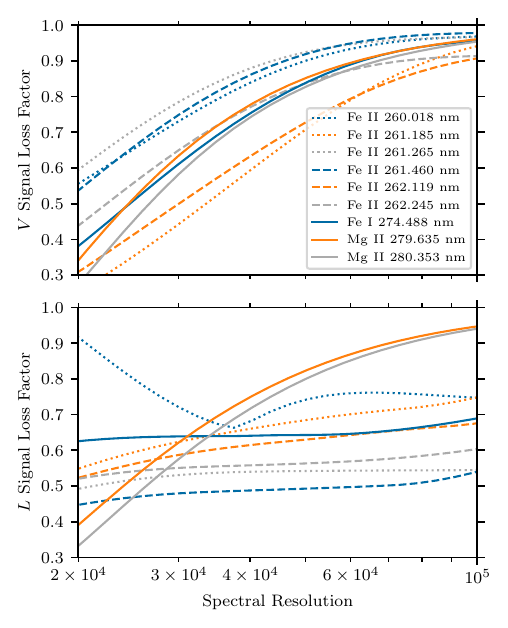}{fig:SNR_loss}{Signal loss factor of SNR resulting from instrumental smearing as a function of the instrument spectral resolution.}

We assume the instrument has a spectral resolution of \num{30000} and calculate $\Gamma'$ and $\mathrm{X}'$ for spectral lines of interest.
A reasonable estimate for a near-optimal, balanced modulator is $\epsilon_I/\epsilon_V\approx\epsilon_I/\epsilon_L\approx1.8$ \citep[e.g.,][]{2010ApOpt..49.3580T}.
We thus set the factors $\epsilon_I/(\sqrt{R}\,\epsilon_V)$ and $\epsilon_I/(\sqrt{R}\,\epsilon_L)$ to be equal to \num{0.01}.
Finally, we have to estimate $B_\bot$ in order to evaluate Eq.~\ref{eq:sigmabbot}.
We choose a stepped function for $B_{\bot}$ starting at \SI{200}{\gauss} at the top of the chromosphere to \SI{50}{\gauss} in the photosphere based on simulations of a magnetic flux rope and a sheared arcade (M.~Rempel, private communication).
As an example, we calculate the required SNR $\sigma_I/I$ required to detect the transverse field and a \SI{20}{\gauss} longitudinal field with $2.5\,\sigma$ significance.
The results are summarized in Table~\ref{tab:SNR}.
We note that the SNR values are the average pixel SNR over the spectral window between the vertical dotted lines in the panels of Fig.~\ref{fig:tblines}.
Generally, we observe that lines that form higher in the atmosphere require higher SNR.
To demonstrate that these SNR requirements are achievable, we also calculate the SNR of a \SI{12}{\s} integration by such a hypothetical instrument with a \SI{30}{\cm} aperture, 2.5\% throughput, and \SI{1}{\arcsec} spatial resolution.
The deep \ion{Fe}{2} line at \SI{260.018}{\nm} drives the instrument requirements.
The \ion{Mg}{2} lines achieve higher SNR than that line because of the increased intensity in the 2V and 2R peaks.

\begin{deluxetable}{RlRR|RR|R}
\tablecaption{\label{tab:SNR}%
Required SNR to detect magnetic field with $2.5\sigma$ significance.
Case 1: detection of transverse field with strength $B_\bot$.
Case 2: detection of $B_\parallel=20~\mathrm{G}$.
The Instrument column lists the projected performance of a hypothetical instrument.
See text for details.}%
\tablehead{%
\colhead{$\lambda_0$} & \colhead{Ion} &
\colhead{$\log\tau_0$} &
\multicolumn{1}{c|}{$B_\bot$} &
\colhead{Case 1} & \multicolumn{1}{c|}{Case 2} &
\colhead{Instrument}}
\startdata
260.018 & \ion{Fe}{2} & \num{-1.39} & \num{150} & \num{539} & \num{419} & \num{ 553} \\
261.185 & \ion{Fe}{2} & \num{-4.23} & \num{50} & \num{185} & \num{127} & \num{ 771} \\
261.265 & \ion{Fe}{2} & \num{-1.79} & \num{150} & \num{263} & \num{228} & \num{ 644} \\
261.460 & \ion{Fe}{2} & \num{-2.20} & \num{100} & \num{351} & \num{184} & \num{ 581} \\
262.119 & \ion{Fe}{2} & \num{-3.75} & \num{75} & \num{ 80} & \num{112} & \num{ 955} \\
262.245 & \ion{Fe}{2} & \num{-2.76} & \num{100} & \num{ 26} & \num{ 65} & \num{ 700} \\
274.488 & \ion{Fe}{1} & \num{-5.51} & \num{50} & \num{108} & \num{ 76} & \num{ 971} \\
279.635 & \ion{Mg}{2} & \num{ 0.00} & \num{200} & \num{818} & \num{768} & \num{1593} \\
280.353 & \ion{Mg}{2} & \num{-0.30} & \num{200} & \num{739} & \num{724} & \num{1436} \\
\enddata
\end{deluxetable}

\section{Conclusion}

We have investigated the lines in the wavelength region between about \num{250} and \SI{290}{\nm} identified by \cite{2021ApJ...917...27J} as being optimal for studying magnetism in the solar chromosphere.
We have derived equations and procedures to quantify the sensitivity of spectral lines to magnetic field through the Zeeman effect based on observed or synthetic intensity spectra.
While we have applied these methods to the spectral region suggested by \cite{2021ApJ...917...27J}, they are applicable to any spectral line.

An example calculation shows that observations with an instrument with a spectral resolution of \num{30000} need to reach achievable Stokes-$I$ SNR ratios of a few hundred, which are achievable for an instrument with a \SI{30}{\cm} aperture and 2.5\% throughput at \SI{1}{\arcsec} spatial resolution with an integration time of \SI{12}{\s}.
We thus conclude that this region of the solar spectrum in the near-UV yields observable polarization signals with suitable diagnostic potential for studies of chromospheric magnetism through the Zeeman effect.

We have not evaluated the \emph{interpretability} of observations of these lines.
That work requires a more complex approach of synthesizing spectra from known model atmospheres, degrading those spectra as if they were observed by a hypothetical instrument, and attempting to recover the model parameters through interpretation using the WFA or with inversion codes like DeSIRe \citep{2022A&A...660A..37R}, STiC \citep{2019A&A...623A..74D}, or TIC \citep{2022ApJ...933..145L}.
The \ion{Mg}{2} lines have been studied and used for diagnostics of chromospheric magnetism in recent years \citep[e.g.,][]{2016ApJ...830L..24D,2019ApJ...883L..30M,2021SciA....7.8406I,2022ApJ...936..115C,2022ApJ...936...67R,2023ApJ...942...60A,2023ApJ...945..144L}
These efforts should be continued and expanded to include the \ion{Fe}{1} and \ion{Fe}{2} lines identified in this work.
\edit1{A first step in this direction was recently taken by \cite{2023ApJ...948...86A}, who studied the magnetic sensitivity of \ion{Fe}{2} between \num{250} and \SI{278}{\nm} using a many-level model atom and realistic physics using the HanleRT code.
They note that observations of the solar spectrum are required to study the effects of UV line blanketing and validate the atomic data, in particular the rate of inelastic collisions with electrons.}
We assert based on the results presented in this paper that this region of the solar spectrum holds great promise and instrumentation to observe it should be developed.

\paragraph{CRediT author statement}
Investigation: A.G.~de Wijn, P.G.~Judge, R.~Ezzeddine, A.~Sainz Dalda;
Writing---original draft preparation: A.G.~de Wijn;
Writing---review and editing: P.G.~Judge.

\begin{acknowledgments}
This material is based upon work supported by the National Center for Atmospheric Research, which is a major facility sponsored by the National Science Foundation under Cooperative Agreement No.\ 1852977.
AdW acknowledges support by the National Aeronautics and Space Administration under Grant 80NSSC21K1792 issued through the Heliophysics Flight Opportunity Studies program.
R.E.\ acknowledges support from NSF grant AST-2206263.
IRIS is a NASA small explorer mission developed and operated by LMSAL with mission operations executed at NASA Ames Research Center and major contributions to downlink communications funded by ESA and the Norwegian Space Centre.
This work has made use of the VALD database, operated at Uppsala University, the Institute of Astronomy RAS in Moscow, and the University of Vienna.
The following acknowledgements were compiled using the Astronomy Acknowledgement Generator (\url{https://astrofrog.github.io/acknowledgment-generator/}).
This research has made use of NASA's Astrophysics Data System,
NumPy \citep{2011CSE....13b..22V},
matplotlib, a Python library for publication quality graphics \citep{2007CSE.....9...90H},
SciPy \citep{2020NatMe..17..261V},
and the IPython package \citep{2007CSE.....9c..21P}.
\end{acknowledgments}

\section*{Comments}
This version of the article has been accepted for publication after peer review but is not the Version of Record and does not reflect post-acceptance improvements, or any corrections. The Version of Record is available online at \url{https://doi.org/10.3847/1538-4357/ace041}.


\begin{thebibliography}{34}
\expandafter\ifx\csname natexlab\endcsname\relax\def\natexlab#1{#1}\fi

\bibitem[{{Afonso Delgado} {et~al.}(2023{\natexlab{a}}){Afonso Delgado}, {del Pino Alem{\'a}n}, \& {Trujillo Bueno}}]{2023ApJ...942...60A}
{Afonso Delgado}, D., {del Pino Alem{\'a}n}, T., \& {Trujillo Bueno}, J. 2023{\natexlab{a}}, \apj, 942, 60

\bibitem[{{Afonso Delgado} {et~al.}(2023{\natexlab{b}}){Afonso Delgado}, {del Pino Alem{\'a}n}, \& {Trujillo Bueno}}]{2023ApJ...948...86A}
{Afonso Delgado}, D., {del Pino Alem{\'a}n}, T., \& {Trujillo Bueno}, J. 2023{\natexlab{b}}, \apj, 948, 86

\bibitem[{{Alvarez} \& {Plez}(1998)}]{1998A&A...330.1109A}
{Alvarez}, R. \& {Plez}, B. 1998, \aap, 330, 1109

\bibitem[{{Casini} \& {de Wijn}(2014)}]{2014JOSAA..31.2002C}
{Casini}, R. \& {de Wijn}, A.~G. 2014, Journal of the Optical Society of America A, 31, 2002

\bibitem[{{Centeno} {et~al.}(2022){Centeno}, {Rempel}, {Casini}, \& {del Pino Alem{\'a}n}}]{2022ApJ...936..115C}
{Centeno}, R., {Rempel}, M., {Casini}, R., \& {del Pino Alem{\'a}n}, T. 2022, \apj, 936, 115

\bibitem[{{de la Cruz Rodr{\'\i}guez} {et~al.}(2019){de la Cruz Rodr{\'\i}guez}, {Leenaarts}, {Danilovic}, \& {Uitenbroek}}]{2019A&A...623A..74D}
{de la Cruz Rodr{\'\i}guez}, J., {Leenaarts}, J., {Danilovic}, S., \& {Uitenbroek}, H. 2019, \aap, 623, A74

\bibitem[{{De Pontieu} {et~al.}(2014){De Pontieu}, {Title}, {Lemen}, {Kushner}, {Akin}, {Allard}, {Berger}, {Boerner}, {Cheung}, {Chou}, {Drake}, {Duncan}, {Freeland}, {Heyman}, {Hoffman}, {Hurlburt}, {Lindgren}, {Mathur}, {Rehse}, {Sabolish}, {Seguin}, {Schrijver}, {Tarbell}, {W{\"u}lser}, {Wolfson}, {Yanari}, {Mudge}, {Nguyen-Phuc}, {Timmons}, {van Bezooijen}, {Weingrod}, {Brookner}, {Butcher}, {Dougherty}, {Eder}, {Knagenhjelm}, {Larsen}, {Mansir}, {Phan}, {Boyle}, {Cheimets}, {DeLuca}, {Golub}, {Gates}, {Hertz}, {McKillop}, {Park}, {Perry}, {Podgorski}, {Reeves}, {Saar}, {Testa}, {Tian}, {Weber}, {Dunn}, {Eccles}, {Jaeggli}, {Kankelborg}, {Mashburn}, {Pust}, {Springer}, {Carvalho}, {Kleint}, {Marmie}, {Mazmanian}, {Pereira}, {Sawyer}, {Strong}, {Worden}, {Carlsson}, {Hansteen}, {Leenaarts}, {Wiesmann}, {Aloise}, {Chu}, {Bush}, {Scherrer}, {Brekke}, {Martinez-Sykora}, {Lites}, {McIntosh}, {Uitenbroek}, {Okamoto}, {Gummin}, {Auker}, {Jerram}, {Pool}, \& {Waltham}}]{2014SoPh..289.2733D}
{De Pontieu}, B., {Title}, A.~M., {Lemen}, J.~R., {et~al.} 2014, \solphys, 289, 2733

\bibitem[{{del Pino Alem{\'a}n} {et~al.}(2016){del Pino Alem{\'a}n}, {Casini}, \& {Manso Sainz}}]{2016ApJ...830L..24D}
{del Pino Alem{\'a}n}, T., {Casini}, R., \& {Manso Sainz}, R. 2016, \apjl, 830, L24

\bibitem[{{del Pino Alem{\'a}n} {et~al.}(2020){del Pino Alem{\'a}n}, {Trujillo Bueno}, {Casini}, \& {Manso Sainz}}]{2020ApJ...891...91D}
{del Pino Alem{\'a}n}, T., {Trujillo Bueno}, J., {Casini}, R., \& {Manso Sainz}, R. 2020, \apj, 891, 91

\bibitem[{{del Toro Iniesta} \& {Collados}(2000)}]{2000ApOpt..39.1637D}
{del Toro Iniesta}, J.~C. \& {Collados}, M. 2000, \ao, 39, 1637

\bibitem[{{Fontenla} {et~al.}(1993){Fontenla}, {Avrett}, \& {Loeser}}]{1993ApJ...406..319F}
{Fontenla}, J.~M., {Avrett}, E.~H., \& {Loeser}, R. 1993, \apj, 406, 319

\bibitem[{{Gustafsson} {et~al.}(2008){Gustafsson}, {Edvardsson}, {Eriksson}, {J{\o}rgensen}, {Nordlund}, \& {Plez}}]{2008A&A...486..951G}
{Gustafsson}, B., {Edvardsson}, B., {Eriksson}, K., {et~al.} 2008, \aap, 486, 951

\bibitem[{{Hall} \& {Anderson}(1991)}]{1991JGR....9612927H}
{Hall}, L.~A. \& {Anderson}, G.~P. 1991, \jgr, 96, 12,927

\bibitem[{{Hunter}(2007)}]{2007CSE.....9...90H}
{Hunter}, J.~D. 2007, Computing in Science and Engineering, 9, 90

\bibitem[{{Ishikawa} {et~al.}(2021){Ishikawa}, {Bueno}, {del Pino Alem{\'a}n}, {Okamoto}, {McKenzie}, {Auch{\`e}re}, {Kano}, {Song}, {Yoshida}, {Rachmeler}, {Kobayashi}, {Hara}, {Kubo}, {Narukage}, {Sakao}, {Shimizu}, {Suematsu}, {Bethge}, {De Pontieu}, {Dalda}, {Vigil}, {Winebarger}, {Ballester}, {Belluzzi}, {{\v{S}}t{\v{e}}p{\'a}n}, {Ramos}, {Carlsson}, \& {Leenaarts}}]{2021SciA....7.8406I}
{Ishikawa}, R., {Bueno}, J.~T., {del Pino Alem{\'a}n}, T., {et~al.} 2021, Science Advances, 7, eabe8406

\bibitem[{{Judge} {et~al.}(2021){Judge}, {Rempel}, {Ezzeddine}, {Kleint}, {Egeland}, {Berdyugina}, {Berger}, {Bryans}, {Burkepile}, {Centeno}, {de Toma}, {Dikpati}, {Fan}, {Gilbert}, \& {Lacatus}}]{2021ApJ...917...27J}
{Judge}, P., {Rempel}, M., {Ezzeddine}, R., {et~al.} 2021, \apj, 917, 27

\bibitem[{{Landi Degl'Innocenti} \& {Landi Degl'Innocenti}(1973)}]{1973SoPh...31..299L}
{Landi Degl'Innocenti}, E. \& {Landi Degl'Innocenti}, M. 1973, \solphys, 31, 299

\bibitem[{{Landi Degl'Innocenti} \& {Landolfi}(2004)}]{2004ASSL..307.....L}
{Landi Degl'Innocenti}, E. \& {Landolfi}, M. 2004, {Polarization in Spectral Lines}, Vol. 307 (Springer, Dordrecht)

\bibitem[{{Li} {et~al.}(2022){Li}, {del Pino Alem{\'a}n}, {Trujillo Bueno}, \& {Casini}}]{2022ApJ...933..145L}
{Li}, H., {del Pino Alem{\'a}n}, T., {Trujillo Bueno}, J., \& {Casini}, R. 2022, \apj, 933, 145

\bibitem[{{Li} {et~al.}(2023){Li}, {del Pino Alem{\'a}n}, {Trujillo Bueno}, {Ishikawa}, {Alsina Ballester}, {McKenzie}, {Auch{\`e}re}, {Kobayashi}, {Okamoto}, {Rachmeler}, \& {Song}}]{2023ApJ...945..144L}
{Li}, H., {del Pino Alem{\'a}n}, T., {Trujillo Bueno}, J., {et~al.} 2023, \apj, 945, 144

\bibitem[{{Manso Sainz} {et~al.}(2019){Manso Sainz}, {del Pino Alem{\'a}n}, {Casini}, \& {McIntosh}}]{2019ApJ...883L..30M}
{Manso Sainz}, R., {del Pino Alem{\'a}n}, T., {Casini}, R., \& {McIntosh}, S. 2019, \apjl, 883, L30

\bibitem[{{Narukage} {et~al.}(2016){Narukage}, {McKenzie}, {Ishikawa}, {Trujillo-Bueno}, {De Pontieu}, {Kubo}, {Ishikawa}, {Kano}, {Suematsu}, {Yoshida}, {Rachmeler}, {Kobayashi}, {Cirtain}, {Winebarger}, {Asensio Ramos}, {del Pino Aleman}, {{\v{S}}t{\k{e}}p{\'a}n}, {Belluzzi}, {Larruquert}, {Auch{\`e}re}, {Leenaarts}, \& {Carlsson}}]{2016SPIE.9905E..08N}
{Narukage}, N., {McKenzie}, D.~E., {Ishikawa}, R., {et~al.} 2016, in Society of Photo-Optical Instrumentation Engineers (SPIE) Conference Series, Vol. 9905, Space Telescopes and Instrumentation 2016: Ultraviolet to Gamma Ray, ed. J.-W.~A. {den Herder}, T.~{Takahashi}, \& M.~{Bautz}, 990508

\bibitem[{{Perez} \& {Granger}(2007)}]{2007CSE.....9c..21P}
{Perez}, F. \& {Granger}, B.~E. 2007, Computing in Science and Engineering, 9, 21

\bibitem[{{Piskunov} {et~al.}(1995){Piskunov}, {Kupka}, {Ryabchikova}, {Weiss}, \& {Jeffery}}]{1995A&AS..112..525P}
{Piskunov}, N.~E., {Kupka}, F., {Ryabchikova}, T.~A., {Weiss}, W.~W., \& {Jeffery}, C.~S. 1995, \aaps, 112, 525

\bibitem[{{Plez}(2012)}]{2012ascl.soft05004P}
{Plez}, B. 2012, {Turbospectrum: Code for spectral synthesis}, Astrophysics Source Code Library, record ascl:1205.004

\bibitem[{{Rachmeler} {et~al.}(2022){Rachmeler}, {Trujillo Bueno}, {McKenzie}, {Ishikawa}, {Auch{\`e}re}, {Kobayashi}, {Kano}, {Okamoto}, {Bethge}, {Song}, {Alsina Ballester}, {Belluzzi}, {del Pino Alem{\'a}n}, {Asensio Ramos}, {Yoshida}, {Shimizu}, {Winebarger}, {Kobelski}, {Vigil}, {De Pontieu}, {Narukage}, {Kubo}, {Sakao}, {Hara}, {Suematsu}, {{\v{S}}t{\v{e}}p{\'a}n}, {Carlsson}, \& {Leenaarts}}]{2022ApJ...936...67R}
{Rachmeler}, L.~A., {Trujillo Bueno}, J., {McKenzie}, D.~E., {et~al.} 2022, \apj, 936, 67

\bibitem[{{Ruiz Cobo} {et~al.}(2022){Ruiz Cobo}, {Quintero Noda}, {Gafeira}, {Uitenbroek}, {Orozco Su{\'a}rez}, \& {P{\'a}ez Ma{\~n}{\'a}}}]{2022A&A...660A..37R}
{Ruiz Cobo}, B., {Quintero Noda}, C., {Gafeira}, R., {et~al.} 2022, \aap, 660, A37

\bibitem[{{Ryabchikova} {et~al.}(2015){Ryabchikova}, {Piskunov}, {Kurucz}, {Stempels}, {Heiter}, {Pakhomov}, \& {Barklem}}]{2015PhyS...90e4005R}
{Ryabchikova}, T., {Piskunov}, N., {Kurucz}, R.~L., {et~al.} 2015, \physscr, 90, 054005

\bibitem[{{Sainz Dalda} {et~al.}(2019){Sainz Dalda}, {de la Cruz Rodr{\'\i}guez}, {De Pontieu}, \& {Go{\v{s}}i{\'c}}}]{2019ApJ...875L..18S}
{Sainz Dalda}, A., {de la Cruz Rodr{\'\i}guez}, J., {De Pontieu}, B., \& {Go{\v{s}}i{\'c}}, M. 2019, \apjl, 875, L18

\bibitem[{{Tomczyk} {et~al.}(2010){Tomczyk}, {Casini}, {de Wijn}, \& {Nelson}}]{2010ApOpt..49.3580T}
{Tomczyk}, S., {Casini}, R., {de Wijn}, A.~G., \& {Nelson}, P.~G. 2010, \ao, 49, 3580

\bibitem[{{Tsuzuki} {et~al.}(2020){Tsuzuki}, {Ishikawa}, {Kano}, {Narukage}, {Song}, {Yoshida}, {Uraguchi}, {Okamoto}, {McKenzie}, {Kobayashi}, {Rachmeler}, {Auchere}, \& {Trujillo Bueno}}]{2020SPIE11444E..6WT}
{Tsuzuki}, T., {Ishikawa}, R., {Kano}, R., {et~al.} 2020, in Society of Photo-Optical Instrumentation Engineers (SPIE) Conference Series, Vol. 11444, Society of Photo-Optical Instrumentation Engineers (SPIE) Conference Series, 114446W

\bibitem[{{Uitenbroek}(2001)}]{2001ApJ...557..389U}
{Uitenbroek}, H. 2001, \apj, 557, 389

\bibitem[{{van der Walt} {et~al.}(2011){van der Walt}, {Colbert}, \& {Varoquaux}}]{2011CSE....13b..22V}
{van der Walt}, S., {Colbert}, S.~C., \& {Varoquaux}, G. 2011, Computing in Science and Engineering, 13, 22

\bibitem[{{Virtanen} {et~al.}(2020){Virtanen}, {Gommers}, {Oliphant}, {Haberland}, {Reddy}, {Cournapeau}, {Burovski}, {Peterson}, {Weckesser}, {Bright}, {van der Walt}, {Brett}, {Wilson}, {Millman}, {Mayorov}, {Nelson}, {Jones}, {Kern}, {Larson}, {Carey}, {Polat}, {Feng}, {Moore}, {VanderPlas}, {Laxalde}, {Perktold}, {Cimrman}, {Henriksen}, {Quintero}, {Harris}, {Archibald}, {Ribeiro}, {Pedregosa}, {van Mulbregt}, \& {SciPy 1. 0 Contributors}}]{2020NatMe..17..261V}
{Virtanen}, P., {Gommers}, R., {Oliphant}, T.~E., {et~al.} 2020, Nature Methods, 17, 261

\end{thebibliography}
\end{document}